\documentclass[aip,pop,onecolumn]{revtex4-1}
\usepackage[final]{graphicx}

\PassOptionsToPackage{hyphens}{url}
\usepackage[hypertexnames=false]{hyperref}
\usepackage{natbib}
\usepackage{amssymb}
\usepackage{amsmath}
\usepackage{color}
\usepackage{xfrac}
\usepackage[normalem]{ulem}

\begin{document}

\title{Speeding Up Simulations By Slowing Down Particles: Speed-Limited Particle-in-Cell Simulation}

\author{Gregory R. Werner}
\affiliation{Center for Integrated Plasma Studies, University of Colorado, Boulder, Colorado 80309, USA}

\author{Thomas G. Jenkins}
\affiliation{Tech-X Corporation, 5621 Arapahoe Avenue Suite A, Boulder, Colorado 80303, USA}

\author{Andrew M. Chap}
\affiliation{Tech-X Corporation, 5621 Arapahoe Avenue Suite A, Boulder, Colorado 80303, USA}

\author{John R. Cary}
\affiliation{Center for Integrated Plasma Studies, University of Colorado, Boulder, Colorado 80309, USA}
\affiliation{Tech-X Corporation, 5621 Arapahoe Avenue Suite A, Boulder, Colorado 80303, USA}


\date{\today}

\begin{abstract}
Based on the particle-in-cell (PIC) plasma simulation method, the speed-limited PIC (SLPIC) method delivers faster kinetic plasma simulation in cases where the particle distributions evolve slowly compared with the maximum stable PIC timestep.
SLPIC thus offers more feasible, fully kinetic simulation in regimes that historically have required fluid approaches, such as magnetohydrodynamic (MHD), two-fluid, or Boltzmann electron treatments.
In particular, SLPIC allows an explicit time advance with steps much larger than the inverse plasma frequency, avoiding the instability explicit PIC faces with large timesteps.
SLPIC avoids this instability by slowing down fast particles (e.g., electrons) in a way that is rigorously underpinned by an approximate Vlasov equation; unlike MHD, two-fluid, and Boltzmann electron approaches, SLPIC does not fundamentally neglect any first-principles plasma physics, although the choices of grid cell size, timestep, and number of macroparticles per cell naturally limit the physical phenomena that can be accurately represented.
SLPIC can be implemented with minor modifications of a standard PIC code and 
does not require an implicit time advance.  It enables large timesteps in first-principles kinetic plasma simulation of appropriately slow phenomena, and it can handle many of the same complications as PIC, such as boundary conditions and collisions.
In an argon plasma sheath test problem, a SLPIC simulation achieved a speed-up of a factor of 160 over the corresponding PIC simulation, without loss of accuracy.
\end{abstract}


\maketitle 

\section{Introduction}

Particle-in-cell (PIC) simulation is a powerful technique for
studying plasma phenomena, in large part because it can include
all of the classical ``first principles'' physics---i.e.,
the Lorentz force and Maxwell's equations (with the latter
sometimes helpfully simplified, e.g., to Poisson's equation).\cite{BirdsallPIC,HockneyPIC}
To accomplish this, PIC simulations track
sample ``macroparticles,'' which follow trajectories of real particles
and which represent discrete realizations of particle distribution functions
in the phase space; statistically, the macroparticle behavior approximates
the distribution function evolution.
PIC simulation is of course still an approximation of the underlying
physics; however, in the limit of infinite grid resolution, infinitesimal
timestep, and small (hence numerous) macroparticles, PIC simulation
includes all the physics of Maxwell's equations and the Lorentz force law.

The very strength of PIC can sometimes be a drawback,
because simulations are often limited
by what one \emph{can} simulate rather
than what one \emph{wants} to simulate.
For example, because explicit electrostatic PIC simulation \emph{can}
capture the phenomenon
of plasma oscillation, such PIC simulations experience instability
unless the timestep $\Delta t$ is small enough to resolve the plasma
frequency.\cite{mason1971computer}
With a timestep determined by the plasma frequency, PIC simulation is
often too costly (computationally) to
simulate phenomena that operate on much longer timescales than plasma
oscillations.

In this manuscript we present  speed-limited PIC (SLPIC), 
a new PIC-based simulation technique
that modifies the PIC method to slow down fast phenomena, enabling larger 
timesteps while retaining the same underlying physics on slow timescales.
SLPIC has the potential to speed up simulations
with fast phenomena that are numerically troublesome, but physically unimportant.
For such simulations, SLPIC improves upon PIC by increasing the amount of approximation; 
however, the degree of approximation can be
continuously varied until SLPIC is identical to PIC, making it possible---without leaving 
the SLPIC framework or sacrificing efficiency---to verify whether the faster phenomena 
in question actually have negligible effect.

Historically, two main classes of methods have been developed to suppress numerical problems related to irrelevant physics: 
reduced-physics methods that neglect (or integrate over) irrelevant physics, and methods that use a special 
time-advance to avoid growth of unresolved plasma modes.  The first class can be very useful when a reduced 
physics model (e.g., magnetohydrodynamics or gyrokinetics) is known and is known to be applicable.
The second class (e.g., 
fully implicit time-advances) can be both difficult to implement and computationally expensive, often involving 
iterative nonlinear solves.  SLPIC does not require a reduced physics model 
and can be explicit; moreover, it can be implemented as a minor modification to an existing PIC code.  
While SLPIC is not a universal replacement 
for either class of methods, it does extend the feasible range of PIC simulations to longer timescales; it has
the potential to be the fastest method in the regime where kinetic simulation is required and distribution 
functions change relatively slowly compared with the plasma frequency.  

Fluid-based simulations are important examples of the first class of methods noted above; 
they use reduced physics models to avoid 
numerical problems with irrelevant phenomena.
Magnetohydrodynamic (MHD)
simulations, for example, lack the ability to simulate plasma oscillations, and thus avoid
any timestep limitation related to the plasma frequency.
Another fluid approach---the use of Boltzmann electrons---treats
electrons as a fluid in thermal equilibrium so that the
electron density is given by $n_0 \exp(e\phi/kT)$, where $\phi$ is the
local electrostatic potential.  The use of Boltzmann electrons permits a PIC (and hence, fully kinetic) treatment
of ions and ion dynamics while neglecting most electron
dynamics.\cite{mason1971computer,cartwright2000nonlinear}
SLPIC may be useful in circumstances where the Boltzmann electron approach is almost sufficient, but more 
physics is required. SLPIC allows the electron distribution to relax to the ion distribution (as Boltzmann 
electrons relax to the potential determined by the ions), but SLPIC can evolve arbitrary electron distributions.
For example, we expect SLPIC to
speed simulation of phenomena such as collisionless sheath formation (in which the electron distribution is not 
Maxwellian) and electron Landau damping of ion-acoustic waves.

Some kinetic PIC-based approaches, such as gyrokinetics,\cite{Lee1983gyrokinetic}
modify the particle equations of motion to integrate over small length and/or time scales
to allow the use of large timesteps.
The SLPIC approach is similar in that it also modifies the particle equations of motion to capture fast phenomena
over large timesteps, but it differs from gyrokinetics because the SLPIC approximation 
does not affect the physics (i.e., does not use a reduced physics model) at slow timescales.
We note that SLPIC could also be applied to a reduced physics model such as gyrokinetics.

In principle, SLPIC is a method for modifying any PIC algorithm (e.g., whether standard Lorentz-force or 
gyrokinetic) by slowing down fast particles.  In SLPIC, slow particles behave just as in normal PIC, while 
fast particles are ``speed-limited''; they locally follow the same phase-space trajectories as real particles, 
but in slow motion.  In the limit of sufficiently slowly-varying fields, SLPIC particles
follow the same phase-space trajectories as real particles (over finite times, not just locally), though at 
different speeds.  It is practical for the speed limit $v_0$ (separating fast and slow particles) to 
depend on the timestep, $v_0 \propto 1/\Delta t$, and thus
SLPIC introduces a continuously-variable approximation (in addition to the PIC approximation) that depends 
on the relative scales of the desired timestep and the temporal variation of particle distribution functions.  
As the timestep decreases below the (stable) PIC timestep, SLPIC simulation becomes identical to PIC simulation.

Although SLPIC can be used with an explicit time-advance, it shares 
some similarities with fully implicit PIC methods (see, e.g., Refs.~\onlinecite{markidis2011energy,chacon2013charge})
that allow larger timesteps without reduced physics or instability.
Naturally and unavoidably, such implicit methods inaccurately simulate phenomena
that are poorly resolved by the timestep; unlike explicit methods, however,
implicit methods may remain stable even when not resolving irrelevant high-frequency
phenomena.  As with implicit methods, the approximation in SLPIC is
continuously adjustable through the choice of timestep.  
The choice of timestep in SLPIC, as in implicit PIC, is a choice to simulate faster
phenomena inaccurately---a choice that is justified when these phenomena are unimportant.

SLPIC has several possible advantages over implicit PIC.  First, SLPIC
can be explicit, and hence faster than implicit PIC (for the same timestep).
Second, SLPIC is very similar to standard PIC (i.e., much simpler than
implicit PIC),
requiring little modification to algorithms except those which govern
individual macroparticle trajectories.  Third, SLPIC handles the problem
of particles crossing too many grid cells within a timestep: crossing
multiple cells in one step leads to inaccuracy and poses a practical challenge for parallel computation.

SLPIC offers a method that (unlike MHD or Boltzmann electrons)
is essentially similar to PIC and does not require any reduced physics
models.  An existing PIC code can be modified
to support SLPIC with relatively little trouble; for example,
field solvers remain completely unchanged.  The main difference is
that (fast) particles move in slow motion.  The most prevalent case where SLPIC 
can speed simulation is perhaps when the electron distribution relaxes (to a quasi-equilibrium) on ion time scales.

SLPIC involves explicit, local modifications of a standard PIC code;
while extra computation is required, that extra computation is
local (involving only an individual particle and its equation of motion) and 
predictable/consistent (i.e., not iterative, or requiring new solvers that might 
affect scaling with problem size or number of parallel processors).

SLPIC is a very new and promising simulation technique.
We introduce and justify the fundamental approach in the
following section.  In subsequent sections, we show calculation of the
plasma frequency in SLPIC and demonstrate the effectiveness of SLPIC
for self-consistent collisionless sheath simulation; we also investigate the ability of SLPIC to simulate resonant wave-particle interaction.

\section{SLPIC}
\label{sec:SLPIC}

The goal of \emph{kinetic} simulation is to find the
self-consistent evolution of the particle
distribution function $f(\mathbf{x},\mathbf{v},t)$, which in a collisionless plasma
can be described by a phase-space continuity equation,
\begin{eqnarray} \label{eg:ConservVlasov}
   \partial_t f(\mathbf{x},\mathbf{v},t)
   + \nabla_x \cdot [\mathbf{v} f(\mathbf{x},\mathbf{v},t)] + \nabla_\mathbf{v} \cdot [\mathbf{a}(\mathbf{x},\mathbf{v},t) f(\mathbf{x},\mathbf{v},t)] = 0,
\end{eqnarray}
where $\mathbf{a}(\mathbf{x},\mathbf{v},t)$ is acceleration due to whatever forces act on a particle located at position $\mathbf{x}$ with velocity $\mathbf{v}$ at time $t$.  (Although we present this
analysis with zero on the right-hand side above, a non-zero value,
e.g., due to collisions, would not alter the SLPIC technique.)
For a Hamiltonian (phase-space-preserving) system,
$\nabla_\mathbf{x} \cdot \mathbf{v} + \nabla_\mathbf{v} \cdot \mathbf{a} = 0$, leading to the familiar Vlasov equation,
\begin{eqnarray} \label{eg:Vlasov}
   \partial_t f(\mathbf{x},\mathbf{v},t)
   + \mathbf{v} \cdot \nabla_\mathbf{x} f(\mathbf{x},\mathbf{v},t) + \mathbf{a} \cdot \nabla_\mathbf{v} f(\mathbf{x},\mathbf{v},t) = 0.
\end{eqnarray}

Equation~(\ref{eg:ConservVlasov})
can be solved by the method of characteristics,
since it describes an element of the phase-space distribution
advecting through phase space with the local velocity and
acceleration.  Thus, it has a solution as a sum over particles
(or trajectories),
\begin{eqnarray} \label{eg:VlasovDeltaFcn}
   f(\mathbf{x},\mathbf{v},t) = \sum_{p} w_p \delta[\mathbf{x} - \mathbf{x}_p(t)] \delta[\mathbf{v} - \mathbf{v}_p(t)],
\end{eqnarray}
where $\mathbf{x}_p$ and $\mathbf{v}_p$ are particle trajectories (and $w_p$ is a weight
representing the number of real particles embodied in macroparticle $p$) -- i.e.,
they satisfy
\begin{subequations}\label{eg:PtclDyn}
\begin{eqnarray} 
   \dot{\mathbf{x}}_p &=& \mathbf{v}_p, \\
   \dot{\mathbf{v}}_p &=& \mathbf{a}(\mathbf{x}_p, \mathbf{v}_p, t).
\end{eqnarray}
\end{subequations}
[One can verify by direct substitution that,
with these equations of motion, Eq.~(\ref{eg:VlasovDeltaFcn})
is a solution of Eq.~(\ref{eg:ConservVlasov}).]
This is the basis of PIC simulation.\cite{BirdsallPIC,HockneyPIC}
PIC methods will not
be reviewed here, except to say that they in essence broaden
the delta function $\delta[\mathbf{x}-\mathbf{x}_p(t)]$ so that particle charges and
currents
can be transferred to a discrete grid for calculation of fields (the scatter
operation), while fields on the grid are interpolated to particle
positions (the gather operation) to yield the forces on the particles.

While this approach can in principle simulate all the fundamental
(classical) physics of plasmas, the separation of scales---especially between
electron and ion motion---often renders practical simulation
impossible with current resources. Two important (and related)
problems make simulation slow: (1) the timestep must generally
be small enough to prevent particles from crossing too many grid cells
in one timestep, and
(2) the timestep must be smaller than the inverse
plasma frequency. The first reason is important for accuracy (as
well as practicality for parallel computing): since fields cannot
vary on length scales smaller than a grid cell, a particle
traversing less than one cell-length in a discrete timestep
experiences only small
changes in fields.  
The second condition is crucial to avoid
catastrophic numerical instability: with a timestep greater than
$2/\omega_{pe}$, where $\omega_{pe}$ is the plasma frequency, the
standard and simple leapfrog integration scheme is unstable,
with numerical solutions that grow exponentially.

If one also wishes to avoid the grid instability (unphysical
heating that increases the Debye length until it is resolved by
the grid), one must usually choose a grid cell length that resolves the electron
Debye length, $\lambda_{De}$, in the case of a stationary, thermal
plasma. In this case, the two criteria above are identical within
a factor of order unity.

For cases where time scales of interest
are long compared with the plasma period, the above conditions on
the timestep are prohibitive.  In steady-state plasma
sheath formation, for example, the relaxation time is 
related to ion time scales,
which are typically longer by $\sim\sqrt{m_i/m_e}$.
For such simulations
we would like to choose a timestep of the order of the
ion plasma period, but we are prevented from doing so by
numerical instability. In such cases, the electrons are
effectively in equilibrium with the electrostatic field.
Critically, they are in a kinetic equilibrium, but a nontrivial
one, since in a collisionless sheath problem, electrons that flow
into the sheath with enough energy to overcome the sheath
potential do not reflect back into the plasma or equilibrate,
so that a Boltzmann dependence $n_e \sim \exp(e \phi/k T_e)$ may not be entirely
accurate.

To address such situations we propose using ``speed-limited''
electrons, which reduce the scale-separation when the dynamics of
interest take place over times that are long compared with the
inverse plasma frequency or the cell-crossing time of the fastest
particles.  To do this we limit the \emph{speed} with which
simulated electrons travel through the simulation to some maximum
$v_0$, but preserve the correct \emph{direction} of travel,
ensuring that a speed-limited electron follows the same path as a real
electron (but at a slower speed). We will show that this approach
allows larger timesteps, and hence faster simulation, while
accurately capturing the physics of longer time
scales---including kinetic effects of electrons on those
timescales.

To use this method, we will simulate a distribution $g(\mathbf{x},\mathbf{v},t)$ of
speed-limited macroparticles, defined through
a speed-limiting function $\beta(\mathbf{x},\mathbf{v},t)$ that we can specify: 
\begin{eqnarray} \label{eg:SpeedLimitVlasovFcn}
   f(\mathbf{x},\mathbf{v},t) = \beta(\mathbf{x},\mathbf{v},t) g(\mathbf{x},\mathbf{v},t).
\end{eqnarray}

Here, $\beta(\mathbf{x},\mathbf{v},t)$ is a function bounded in the range
(0,1], and $f(\mathbf{x},\mathbf{v},t)$ is the conventional (physical)
distribution function.
PIC evolves $f(\mathbf{x},\mathbf{v},t)$ by moving a collection of macroparticles
along phase-space-preserving trajectories, and we will formulate
SLPIC to evolve $g(\mathbf{x},\mathbf{v},t)$ in a similar manner, noting that 
$g(\mathbf{x},\mathbf{v},t)$ can be trivially converted to $f(\mathbf{x},\mathbf{v},t)$ at any time. 
Inserting the representation of Eq.~(\ref{eg:SpeedLimitVlasovFcn}) 
into Eq.~(\ref{eg:ConservVlasov}) yields
\begin{eqnarray}
   \partial_t [\beta g(\mathbf{x},\mathbf{v},t)]
   + \nabla_\mathbf{x} \cdot [\beta \mathbf{v} g(\mathbf{x},\mathbf{v},t)] +
   \nabla_\mathbf{v} \cdot [\beta \mathbf{a}(\mathbf{x},\mathbf{v},t) g(\mathbf{x},\mathbf{v},t)] = 0,
\end{eqnarray}
which we rewrite in the form
\begin{eqnarray} \label{eg:ConservSpeedLimitVlasov}
   \partial_t g(\mathbf{x},\mathbf{v},t)
   + \nabla_\mathbf{x} \cdot [\beta \mathbf{v} g(\mathbf{x},\mathbf{v},t)] +
   \nabla_\mathbf{v} \cdot [\beta \mathbf{a}(\mathbf{x},\mathbf{v},t) g(\mathbf{x},\mathbf{v},t)] =
   \partial_t [(1 -\beta) g(\mathbf{x},\mathbf{v},t)],
\end{eqnarray}
The approximation that makes SLPIC useful, and which we shall adopt hereafter, 
is the neglect of the
right-hand side of Eq.~(\ref{eg:ConservSpeedLimitVlasov}).  With this
approximation, the equation may [in the same manner as Eq.~(\ref{eg:Vlasov})] be
solved using the method of characteristics, though we will later show that the characteristic
phase-space trajectories differ from characteristic Vlasov trajectories in important and useful ways.  
In the limit that this right-hand side vanishes
exactly, SLPIC is as accurate as PIC.
In other words, SLPIC achieves full-PIC accuracy
for steady-state scenarios, where the right-hand side of Eq.~(\ref{eg:ConservSpeedLimitVlasov}) vanishes.
Also, when $\beta = 1$, the right-hand side vanishes even when
steady-state has not been reached, and again SLPIC achieves exactly the same 
accuracy as standard PIC.  

Effective use of SLPIC involves choosing $\beta(\mathbf{x},\mathbf{v},t)$ and hence the imposed 
speed limit $v_0$.  
For performance (i.e., to allow large timesteps), we want $v_0$ to be as low as possible.
For accuracy, however, the speed limit must be high enough not to interfere with the kinetic behaviors of interest
(e.g., in a simple case where electron kinetics are known to be unimportant, the speed limit could be set just above the speed of the
fastest ions, but below electron velocities).
To illustrate, we consider (not necessarily small) perturbations
or plasma modes with characteristic phase velocities $v_\phi$. Particles moving much faster than
such waves equilibrate with them rapidly, and so $f(\mathbf{x},\mathbf{v},t)$ is quasi steady-state and we can neglect its time derivative. On the other hand,
particles with velocities near $v_\phi$ can interact strongly with the waves, e.g., via Landau damping, and thus may not be in equilibrium with the perturbations.
For such particles all temporal derivatives in Eq.~(\ref{eg:ConservSpeedLimitVlasov})
must be kept to model the physics correctly; if we set $\beta$ to be unity or nearly
unity for velocities less than $v_0$ and set $v_0$ such that $v_0 \gtrsim v_\phi$,
then the right side of Eq.~(\ref{eg:ConservSpeedLimitVlasov})
vanishes for all slow particles, including the strongly-interacting particles (see 
\S\ref{sec:waveParticle} for more exploration of wave-particle interactions). Thus, it is a uniform approximation to set the
right-hand side of Eq.~(\ref{eg:ConservSpeedLimitVlasov}) to zero:
\begin{eqnarray} \label{eq:ConservSpeedLimApprox}
   \partial_t g(\mathbf{x},\mathbf{v},t)
   + \nabla_\mathbf{x} \cdot [\beta \mathbf{v} g(\mathbf{x},\mathbf{v},t)] +
   \nabla_\mathbf{v} \cdot [\beta \mathbf{a}(\mathbf{x},\mathbf{v},t) g(\mathbf{x},\mathbf{v},t)] = 0,
\end{eqnarray}
since it is valid for high velocities by accurately giving their
equilibrium, and it is valid for low velocities because $\beta\approx 1$.

With this approximation, we can evolve
$g(\mathbf{x},\mathbf{v},t)$ according to
\begin{eqnarray} \label{eg:SpeedLimDeltaFcn}
   g(\mathbf{x},\mathbf{v},t) &=& \sum_p w_{p} \delta[\mathbf{x} - \mathbf{x}_{p}(t)] \delta[\mathbf{v} - \mathbf{v}_{p}(t)]
\end{eqnarray}
by the method of characteristics, just as is done for Eq.~(\ref{eg:ConservVlasov}).
To satisfy Eq.~(\ref{eq:ConservSpeedLimApprox}) the
macroparticles must evolve along slow (or speed-limited)
trajectories $\mathbf{x}_p(t)$, $\mathbf{v}_p(t)$ satisfying
the equations of motion 
\begin{subequations}\label{eg:LimPtclDyn}
\begin{eqnarray} 
   \dot{\mathbf{x}}_{p} &=& \beta(\mathbf{x}_{p},\mathbf{v}_{p},t) \mathbf{v}_{p}, \\
   \dot{\mathbf{v}}_{p} &=& \beta(\mathbf{x}_{p},\mathbf{v}_{p},t) \mathbf{a}(x_{p}, \mathbf{v}_{p}, t).
\end{eqnarray}
\end{subequations}
When $\beta = 1$, macroparticles are evolved in the same manner as in conventional PIC, while
for $\beta < 1$, macroparticles are evolved more slowly along their trajectories, with simulation velocity $\dot{\bf x}_p = \beta \mathbf{v}_p$ instead of $\mathbf{v}_p$.
The motion of SLPIC macroparticles approximates the time-evolution 
of $g(\mathbf{x},\mathbf{v},t)$---and this is computationally faster than the analogous
tracking of PIC macroparticles to approximate the time-evolution of $f(\mathbf{x},\mathbf{v},t)$, 
since even the fastest SLPIC macroparticles move slowly enough to enable
the use of much larger timesteps than PIC permits.
It is important to point out that the SLPIC distribution $g(\mathbf{x},\mathbf{v},t)$ is not the
same as the physical distribution function $f(\mathbf{x},\mathbf{v},t)$; it obeys a different
kinetic equation with different phase-space characteristics.  
Nevertheless,
slow physics processes from $f(\mathbf{x},\mathbf{v},t)$ evolution can be replicated as
$g(\mathbf{x},\mathbf{v},t)$ evolves, for a suitably chosen speed-limiting function $\beta$.
Further, with this known $\beta(\mathbf{x},\mathbf{v},t)$, one can convert $g$ to $f$ as desired.

In some sense, $\beta$ acts like a macroparticle
weight; i.e., a macroparticle that is used to evolve $g$ is
subsequently ``weighted'' by $\beta(\mathbf{x},\mathbf{v},t)$ to compute
$f$.  With this view, $f$ is the sum of macroparticles following
trajectories given by Eqs.~(\ref{eg:LimPtclDyn}) with weights changing
according to $(d/dt) \beta$ evaluated along each particle's trajectory.

There are many possibilities for the speed-limiting function
$\beta(\mathbf{x},\mathbf{v},t)$.  For SLPIC,
$\beta$ needs to limit the speed of macroparticles to some value $v_0$.
For large velocities, we must have $\beta \approx v_0/|\mathbf{v}|$ (to limit
$|\dot{\mathbf{x}}|$ to $v_0$), and
for small velocities, $\beta \approx 1$ [so that these macroparticles
evolve according to conventional PIC dynamics].   Examples are
\begin{subequations}
\begin{eqnarray}
   \beta &=& \frac{v_0}{\sqrt{\left|\mathbf{v}\right|^2 + v_0^2}} \label{eq:relativisticBeta} 
  \\
  \label{eq:heavisideBeta}
   \beta &=& \Theta(v_0 - \left|\mathbf{v}\right|) + \frac{v_0}{\left|\mathbf{v}\right|} \Theta(\left|\mathbf{v}\right|-v_0)
\end{eqnarray}
\end{subequations}
\begin{figure}
   \includegraphics{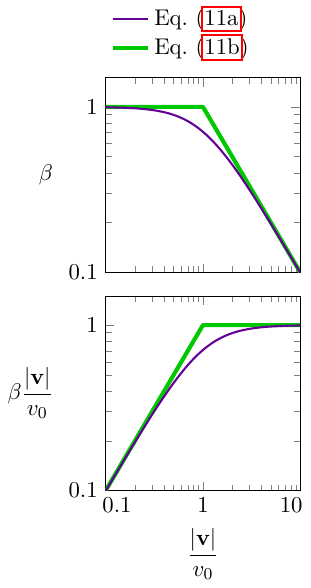}
   \caption{\label{fig:beta}\textit{Top:} Example functions for $\beta$ as a function of particle velocity normalized by the speed limit.  \textit{Bottom:} For those same functions, the speed at which a particle moves through a SLPIC simulation as a function of its physical velocity, both normalized by the speed limit.}
\end{figure}
(see Fig.~\ref{fig:beta} for plots of these functions) where $\Theta$ is the Heaviside step function,
but there are many other options.  It
is possible for $\beta$ to have spatial dependence, e.g.,
for a spatially varying grid, or time-dependence, e.g, to adjust
the severity of approximation mid-simulation; simpler $\beta$
functions that depend only on velocity are also useful for many applications.
In any case, we will denote the speed
limit by $v_0$, keeping in mind that, in principle, it may vary with position and time.

An important aspect of this method is that, since the right-hand sides
of Eqs.~(\ref{eg:PtclDyn}) and (\ref{eg:LimPtclDyn}) differ by a
scalar factor $\beta$,
speed-limited particles [representing $g(\mathbf{x},\mathbf{v},t)$] follow (locally, given the same fields/forces)
the same phase-space trajectories as real particles [representing
$f(\mathbf{x},\mathbf{v},t)$], except at a slower speed. Therefore, even though
fast electrons (with $\left|\mathbf{v}\right| \gg v_0$) move
unphysically slowly (with $\left|\dot{\mathbf{x}}\right|\sim v_0$) under SLPIC,
they follow the correct path
as long as the fields evolve slowly. As a speed-limited particle (with
$\left|\mathbf{v}\right| \gg v_0$) accelerates, its actual speed $\left|\mathbf{v}\right|$ increases
(so it gains energy),
but $\left|\dot{\mathbf{x}}\right|$ remains near $v_0$; to compensate,
its weight must decrease.  E.g., in a steady-state streaming
fluid, an increase in velocity results in a decrease in density; in SLPIC,
the macroparticle speed doesn't change much, so the macroparticle density doesn't change much, and the macroparticle weight
decreases to reflect the real decrease in density.

With particles limited to speeds below $v_0$, one may
choose the timestep $\Delta t \lesssim \Delta x/v_0$, where
$\Delta x$ is the cell size, so that particles will not cross
more than one cell per timestep.  This allows an increase in timestep
by a factor $v_{\rm max}/v_0 \sim v_{te}/v_0$, where $v_{\rm max}$
is the maximum particle speed and $v_{te}$ is the thermal velocity.
Like methods involving Boltzmann
electrons, this method is useful when the electron distribution
is quasi steady-state on the timescales of ion motion.  However,
unlike Boltzmann electron methods, this method simulates an arbitrary electron distribution and has an ``adjustable''
approximation, which allows the simulation
to change continuously into a full PIC simulation by increasing
$v_0$ above the maximum particle speed.

Despite some mathematical resemblance,
SLPIC is not a $\delta \! f$-PIC method,\cite{Parker1993fully, Hu1994}
nor is it equivalent to lowering the speed of light to $v_0$.
Whereas $\delta \! f$ methods evolve a perturbed distribution
$\delta \! f(\mathbf{x},\mathbf{v},t)$ on top of a given (usually equilibrium) distribution
$f_0(\mathbf{x},\mathbf{v})$,
SLPIC evolves a distribution $g(\mathbf{x},\mathbf{v},t)$ which reproduces key physics
processes occurring in the entire distribution $f(\mathbf{x},\mathbf{v},t)$; 
it does not require that the solution be a small perturbation of some
known equilibrium.  And while lowering the speed of light to $v_0$ would certainly impose a speed limit, it would also alter particle trajectories in a way that SLPIC doesn't.

\section{Plasma oscillations for speed-limited electrons}

As was noted above, using speed-limited electrons allows us
to relax the cell-crossing timestep restriction, as electrons
move more slowly through the simulation.
It turns out that speed-limiting of
electrons also lowers the electron plasma frequency, which
relaxes the other condition that required a small timestep. Here
we show that the plasma frequency for speed-limited electrons is
reduced by $\sim v_0/v_{te}$
(again, allowing the timestep to be increased by a
factor of $v_{te}/v_0$, where $v_{te}$ is the electron thermal velocity).

To compute the plasma frequency in the SLPIC system,
we consider 1D wave-like
perturbations $\exp[i(kx-\omega t)]$ from a zero-field, uniform,
steady-state distribution $g_0$ with $\beta=\beta(v)$ independent of
space and time.
Denoting unperturbed quantities with subscript 0, and
first-order with subscript 1, the first-order solution to
Eq.~(\ref{eq:ConservSpeedLimApprox}) for the speed-limited
distribution function is
\begin{eqnarray} \label{eq:speedlimpert}
  -i(\omega -k \beta v) \tilde{g_1} =
  - \tilde{a_1} \partial_v [\beta g_0(v)],
\end{eqnarray}
where $a_1$ is the acceleration due to the first-order electric field,
and tildes indicate amplitudes of oscillation, i.e.,
\begin{eqnarray}
  g_1(x,v,t) = \tilde{g_1}(v)\exp(ikx - i \omega t).
\end{eqnarray}
From this we find the solution for the density perturbation:
\begin{eqnarray}
  \tilde{n_1} &=& \int dv \, \tilde{f_1} = \int dv \, \beta \tilde{g_1}
             \;=\; -i \tilde{a_1} \int dv \,  \frac{\beta}{\omega - k \beta v}
                \partial_v [\beta g_0] \nonumber \\
             &=& i \tilde{a_1} \int dv \, \beta g_0
                 \partial_v \frac{\beta}{\omega - k \beta v}
             \;=\; i \tilde{a_1} \int dv \, f_0
                 \partial_v \frac{\beta}{\omega- k \beta v} \nonumber \\
             &=& i \tilde{a_1} n_0 \left\langle \partial_v
                 \frac{\beta}{\omega- k \beta v} \right\rangle
,\end{eqnarray}
where the angled brackets denote the average over the
velocity distribution function $f_0$. Inserting this into Gauss's
law [$\partial_x E_1 = (-e)n_1/\epsilon_0$ or
$i k \tilde{a_1} = \tilde{n_1} e^2/(m\epsilon_0)$] we find
\begin{eqnarray}
  1 & = & \frac{\omega_p^2}{k} \left\langle \partial_v
                 \frac{\beta}{\omega - k \beta v} \right\rangle
    \; = \; \frac{\omega_p^2}{\omega^2} \left\langle
                 \frac{\beta^2 + (\partial_v \beta) \omega/k}
                 {(1 - k \beta v/\omega)^2} \right\rangle
,\end{eqnarray}
where $\omega_p^2 = e^2 n_0/(m\epsilon_0)$.  The plasma frequency
is found by looking at the long wavelength limit
$k \rightarrow 0$.
When $\beta=1$ (hence $\partial_v \beta = 0$), we recover the
standard result: $\omega=\omega_p$.

When $\partial_v \beta \neq 0$, then for isotropic $f_0$ and $\beta$
[i.e., $f_0(-v)=f_0(v)$ and $\beta(v)=\beta(-v)$],
$\int dv f_0(v) \partial_v \beta(v) = 0$, and expansion of the
denominator to first order yields
\begin{eqnarray}
  \omega^2 = \omega_p^2 \left\langle \beta^2 + 2 \beta \beta' v \right\rangle
.\end{eqnarray}
When the limiting velocity $v_0$ is much less than the
electron thermal velocity $v_{te}$, the speed-limiting function
is $\beta \sim v_0/|v|$ and
the speed-limited plasma frequency $\omega_s$ is
\begin{eqnarray}
  \omega_s \sim \omega_p \frac{v_0}{v_{te}}
.\end{eqnarray}
I.e., the effective plasma frequency is reduced by nearly the same fraction
by which a typical particle's speed is limited, $v_0/v_{te}$.

By speed-limiting simulated particles to $v_0$, i.e., reducing the
speed of typical electrons by $v_0/v_{te}$, we reduce the
simulated plasma frequency by a similar factor.  Naturally,
plasma oscillations are not accurately simulated, but that is an
advantage because this method is appropriate only for cases where the
important dynamics are much slower than plasma oscillations.

\section{SLPIC for a 1D plasma sheath}\label{sec:sheath}

To demonstrate its basic usefulness and accuracy, we implemented
a speed-limited PIC simulation algorithm in VSim (a software package containing the Vorpal computational engine \cite{Nieter:2004}), 
an electro\-magnetic/electro\-static
particle-in-cell/finite-difference time-domain code, applying it to
an electrostatic collisionless sheath simulation with argon plasma ($m_i/m_e=39.9\times 1836$) with one spatial dimension but 3-dimensional velocities and field vectors (1D3V).
Maintaining a constant $\Delta \phi = 12.5$~V potential difference
between conducting plates separated by $L_x = 4$~cm, we inject electrons and argon ions from the right side; ions accelerate toward the left wall, where they are absorbed; high-energy electrons, with energy sufficient to overcome the 12.5~V barrier, are absorbed by the left wall, while lower-energy electrons reflect and are absorbed by the right wall.
Ions and electrons are injected from the right side as if from a stationary plasma of density $5.0\times 10^{13}~{\rm m}^{-3}$ with electron temperature $T_e=2.2$~eV ($v_{te}=6.2\times 10^5$~m/s) and ion temperature $T_i=0.5$~eV ($v_{ti}=1100$~m/s)---with zero mean velocity (e.g., ions do not enter with any bulk drift speed; as we later show, an injection sheath forms that accelerates ions approximately to the Bohm velocity).

\begin{figure}
   \centering
   \includegraphics*{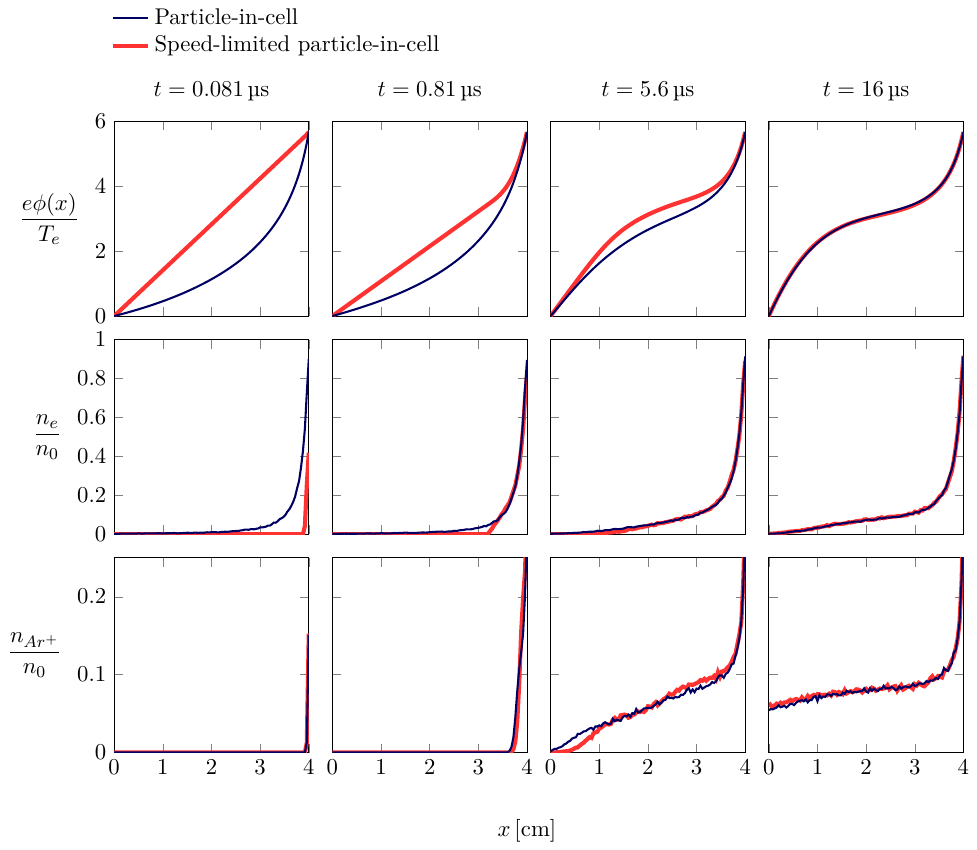}
   \caption{
   \label{fig:SLPICvsPIC}
   Normalized electric potential $\phi(x)$, electron density $n_e(x)$, and ion density $n_{Ar+}(x)$ at various times in PIC and SLPIC simulations.
   \textit{Column 1, $t=2\Delta t_{\rm SLPIC}=1.3L_x/v_{te}$:} Only the PIC electrons have crossed the gap, while ions and SLPIC electrons have hardly moved; the SLPIC potential is nearly unchanged from the initial condition, differing greatly from the PIC potential that reflects the negative space charge in the simulation.
   \textit{Column 2, $t=13L_x/v_{te}=0.1t_{\rm cross}$:} Ions and SLPIC electrons have traveled less than 1~cm into the simulation, and PIC and SLPIC potentials have become similar in that range.
   \textit{Column 3, $t=0.7t_{\rm cross}=1.1L_x/v_0$:} SLPIC electrons and faster ions have crossed the gap; SLPIC and PIC densities are somewhat similar, and potentials are nearing the steady state.
   \textit{Column 4, $t=2t_{\rm cross}$:} After 2 ion-crossing times, PIC and SLPIC agree closely, having almost reached steady state.
The SLPIC simulation ran $160$ times faster than the PIC simulation.
      }
\end{figure}

We simulated the sheath with $N_x=100$ cells and cell size $\Delta x=N_x/L_x$, giving about 4 cells per Debye length (0.16~cm in the uniform plasma).
We then chose the timestep so that (for normal PIC simulation) 
the fastest electrons cross less than one cell per timestep: $\Delta t_{\rm PIC} = \Delta x / (5 v_{te}) = 1.3\times 10^{-10}$~s (we note that since the electrons have three Maxwellian velocity components,
each with standard deviation $v_{te}$, 3\% of all particles have scalar
velocities exceeding $3 v_{te}$ but less than 0.002\% have scalar
velocities exceeding $5 v_{te}$, hence the factor of 5 in the preceding equation).  
Although the timestep is determined by the fastest electrons, the
simulation approaches a steady state after a few ion-crossing times;
the ratio between these scales is large and proportional to the ion/electron scale ratio $\sqrt{m_i/m_e}\approx 270$: 
\begin{equation}
  {t_{\text{cross}} \over \Delta t} \sim \frac{L_x/v_{d}}{\Delta x / (5 v_{te})} \sim
  5N_{x} \sqrt{{m_{i} \over m_{e}}} \sqrt{{T_{e} \over q_{i} \Delta \phi}}
  \sim 6 \times 10^4
\end{equation}
where $v_d$ is a typical ion velocity 
(e.g., $m_i v_d^2/2 \sim e\Delta \phi/2$ or $v_d\sim 0.5\:$cm/$\mu$s, 
hence $t_{\rm cross}\sim 8\:\mu$s).
The simulation reaches equilibrium in roughly $24\:\mu$s, or 3 ion crossing times, or $1.9\times 10^5$ timesteps.
In both PIC and SLPIC simulations, the number of macroparticles varies with time, settling down to approximately $8.3\times 10^4$ ions and $1.1\times 10^5$ electrons in the final steady state, roughly $10^3$ macroparticles per cell on average, though cells on the left side have many fewer macroparticles.

On the other hand, with SLPIC simulation, we can
increase the timestep by $\sim \sqrt{m_i/m_e}$, allowing much faster simulation.
Using $\beta$ from Eq.~(\ref{eq:heavisideBeta}), we
set the SLPIC speed limit to $v_{0} = 9.8 \times 10^{3}$~m/s~$=0.016 v_{te}$, which is above the speed of all ions in the simulation at steady-state (an ion would need 20~eV to have $v\approx v_0$, so the ions behave exactly as in normal PIC).
Again choosing the timestep so that (speed-limited) electrons cannot cross more than one cell per step, we set $\Delta t_{\rm SLPIC} = 320 \Delta t_{\rm PIC} = \Delta x/v_0$. 
The SLPIC simulation reaches the same 
steady state as PIC after about 600 timesteps (or 24$\:\mu$s) simulated time, as in PIC), approximately 160 
times faster 
than the PIC simulation. (We note that the speedup observed here is
a factor of two less than the timestep increase $\Delta t_{\rm SLPIC}/\Delta t_{\rm PIC}=320$, due to the additional
computational work required by the SLPIC particle push; this is discussed in the appendix.)
As in PIC, the simulation is quite near equilibrium even after $16\:\mu$s or 2 ion crossing times.

Figure~\ref{fig:SLPICvsPIC} demonstrates that SLPIC yields the same results as PIC as steady-state is approached.  
In the first column (at $t=2\Delta t_{\rm SLPIC}=1.3L_x/v_{te}$), SLPIC particles and PIC ions have traveled at most 2 cells, while PIC electrons have already crossed the entire simulation; thus the SLPIC potential still resembles its initial state, while the negative space charge in the PIC simulation creates the concave-up potential.  Most of the PIC electrons, however, are reflected before traveling more than $\simeq 1$~cm due to the potential drop of $3T_e/e$.
In the second column ($t=20\Delta t_{\rm SLPIC}=13L_x/v_{te}$), PIC ions and SLPIC particles have traveled almost 1~cm from injection, and the PIC and SLPIC potentials are somewhat similar in this range, though still different for $x \lesssim 3\:$cm.
The third column ($t\sim 0.7 t_{\rm cross}=1.1L_x/v_0$) shows the simulation a little after the first time that many PIC ions and SLPIC particles have crossed the simulation; the potential is nearing its steady state value (in column 4), though neither PIC nor SLPIC has converged.  At this time, the electron density (in both PIC and SLPIC) is also near equilibrium, but slower ions have not yet had time to cross the gap, and so the ion density at $x\lesssim 2$~cm is low compared with the steady-state density.  The ion densities in PIC and SLPIC are by this time very similar; although PIC and SLPIC time evolution differed significantly at earlier times, they now evolve similarly, because the time evolution is governed by the ions, which follow the same trajectories in PIC and SLPIC.
In the fourth column ($t=2t_{\rm cross} = 3.1L_x/v_0$), PIC and SLPIC show essentially identical potentials and particle densities very close to equilibrium (they would be indistinguishable from equilibrium values on the scale of the graphs shown).
It's interesting to note that an injection sheath forms that accelerates ions to $\simeq 2.5\:$V, approximately the Bohm velocity~$\sqrt{T_e/m_i}$.

\section{Wave-particle interactions}
\label{sec:waveParticle}
A fundamental class of plasma physics interactions between particles and waves---e.g., (inverse) Landau damping---depends critically on particles traveling with their correct velocities.
As noted in \S\ref{sec:SLPIC}, only particles with velocities near a wave phase velocity $v_\phi$ interact strongly with the wave fields; these particles are ``resonant'' with the wave---that is, they travel with or ``surf'' the wave.
The use of SLPIC with $v_0 \lesssim v_\phi$, slowing down particles that could otherwise resonate with the wave, is inappropriate when resonant wave-particle interaction is important.
If $v_\phi$ is near the speed of the fastest particles, setting $v_0 \gg v_\phi$ would make SLPIC essentially identical to PIC, offering no advantage.
However, SLPIC may simulate wave-particle interaction to advantage when $v_\phi$ is slower than the fastest particles (such as for Landau damping of ion acoustic waves).
In this section, we investigate a lower bound on $v_0/v_\phi$ necessary for accurate representation of resonant particle-wave interaction
by considering a 1D example of SLPIC test particles in a prescribed electric field resembling a Langmuir wave.

\begin{figure}
   \centering
   \includegraphics[width=.90\textwidth]{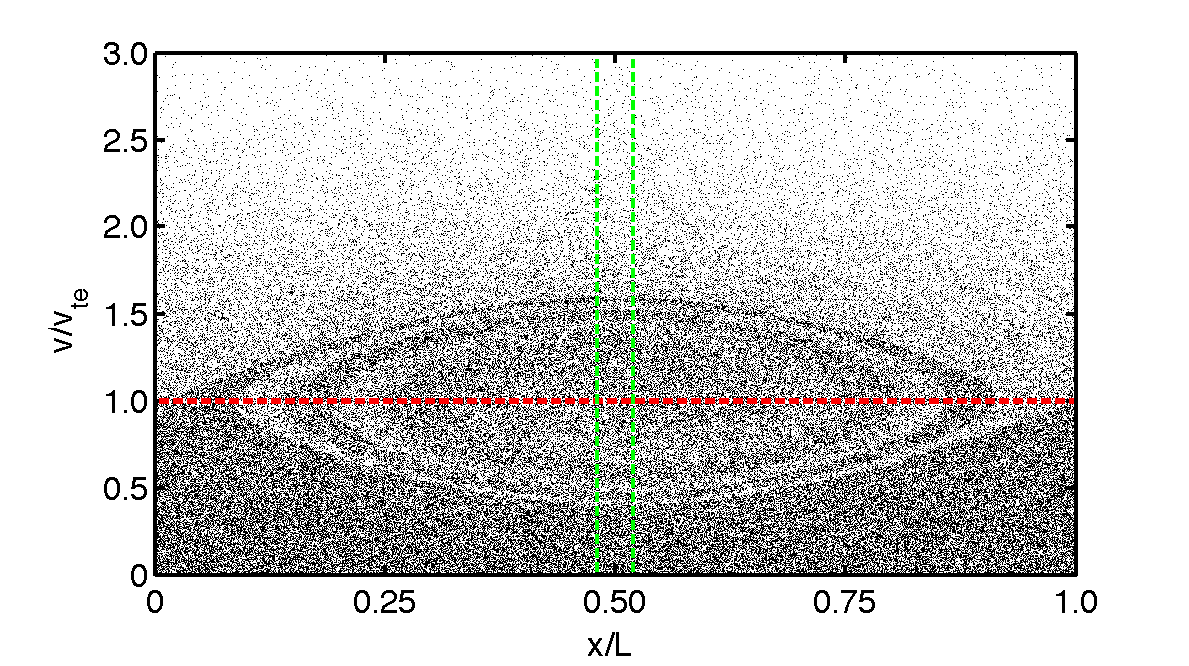}
   \caption{
   \label{fig:resonantPIC}
   As expected, PIC accurately simulates
   the trapping of initially-Maxwellian electrons in 1D phase space arising from a prescribed electrostatic wave $\phi(x,t) = \phi_{0} \sin \left[2\pi(x-v_{te}t)/L\right] $ in a periodic box with phase velocity $v_\phi=v_{te}$ (horizontal red line).  
   Electrons trapped by the wave fields reside within the separatrix (eye-like phase space structure).  The particle distribution within the vertical
green lines is shown in Fig.~\ref{fig:DistPlot}.
}
\end{figure}

We start with a 1D non-drifting Maxwellian velocity distribution of electrons at temperature
$T_{e} = 1.0\:$eV (i.e., $v_y=v_z=0$) in a 1D periodic domain of length $L = 4.0\:$cm.  
We impose a traveling wave in the electric potential of the form $\phi(x,t) = \phi_{0}(t) \sin \left[k(x-v_{te}t)\right] $, with wavenumber $k=2\pi/L$ and frequency $\omega = k v_{te}$ chosen so that $v_\phi=\omega/k=v_{te}$.  The wave amplitude $\phi_{0}(t)$ is slowly increased from $0\:$V to $0.1\:$V over twenty wave periods and remains constant thereafter; $\phi(x,t)$ is unaffected by the particles.
PIC simulation (Fig.~\ref{fig:resonantPIC}) correctly shows that,
over time,
a localized quasilinear flattening of the electron distribution function occurs about the resonant velocity due to wave-particle resonance (or ``trapping''---particles slower than the wave tend to be accelerated, while faster particles decelerate).
The saturated width of the flattened region (the vertical extent of the eye-like shape in Fig.~\ref{fig:resonantPIC}) approaches a constant value in time as the particle distribution relaxes to a new quasi-steady-state configuration in the presence of the traveling wave; this width in velocity space is proportional to $\lim_{t \rightarrow \infty}\sqrt{e \phi_{0}(t)/T_{e}}$.

\begin{figure}
   \centering
   \begin{tabular}{@{}l@{}l@{}}
\includegraphics*[width=.49\textwidth,trim=0.34in 0.06in 0.6in 0.28in]{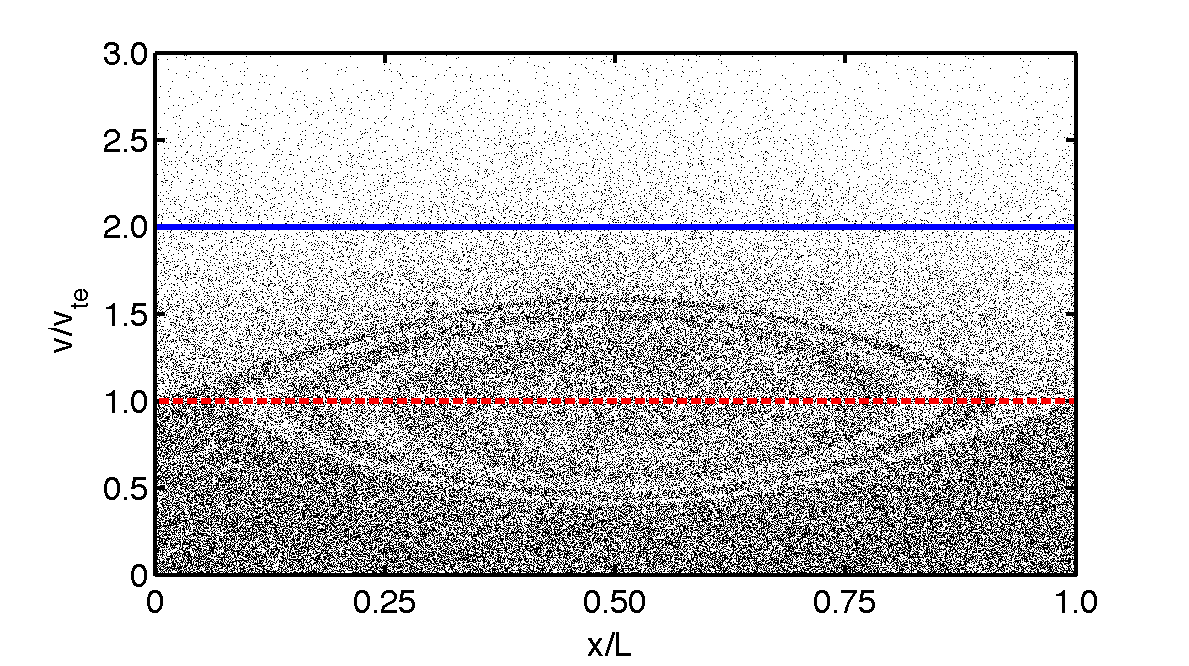}&%
\includegraphics*[width=.49\textwidth,trim=0.34in 0.06in 0.62in 0.28in]{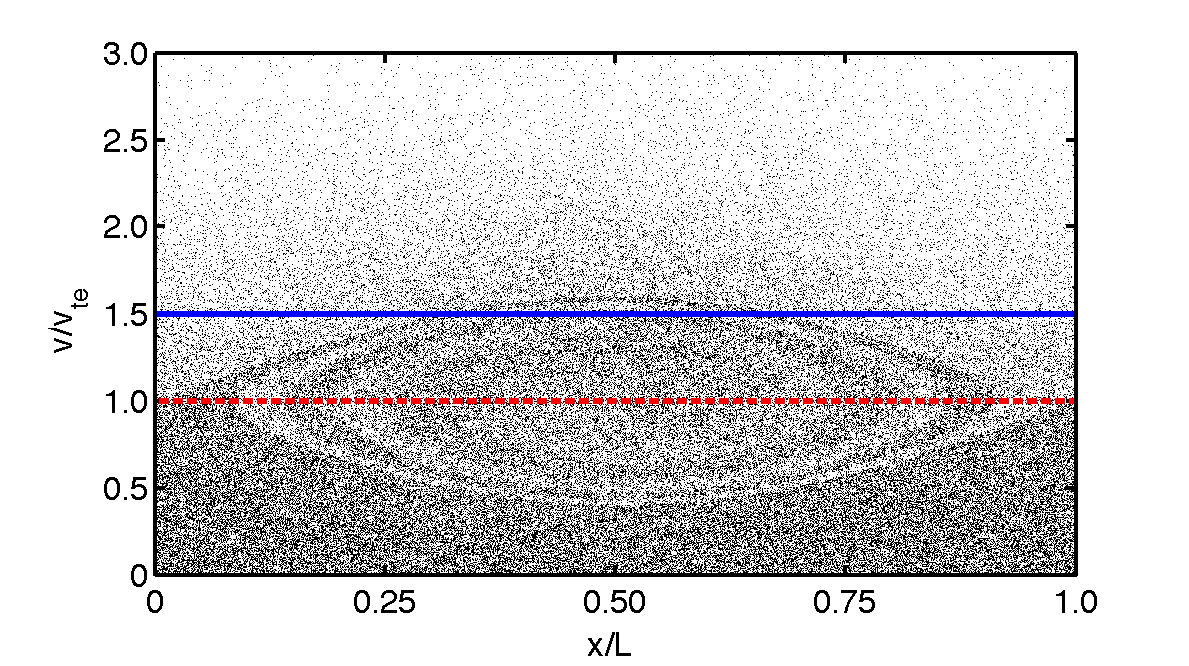}\vspace{-0.3in}\\[0in]
(a) $v_0=2v_\phi$ &(b) $v_0=1.5v_\phi$ \\[0.1in]
\includegraphics*[width=.49\textwidth,trim=0.34in 0.06in 0.6in 0.28in]{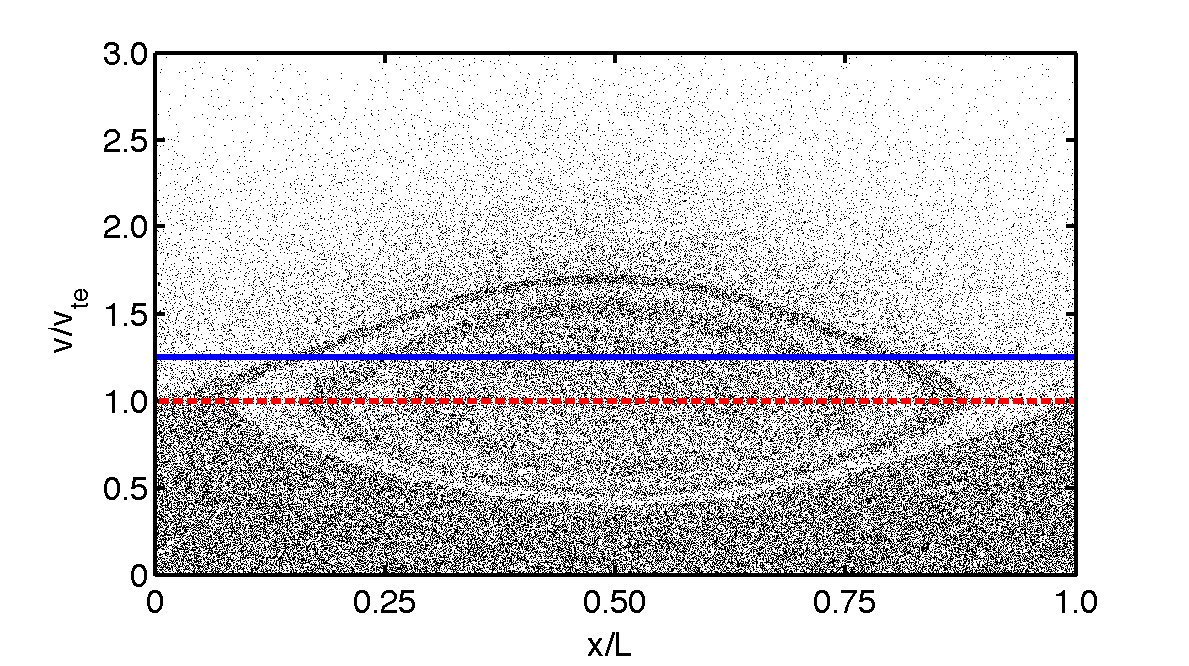}&%
\includegraphics*[width=.49\textwidth,trim=0.34in 0.06in 0.62in 0.28in]{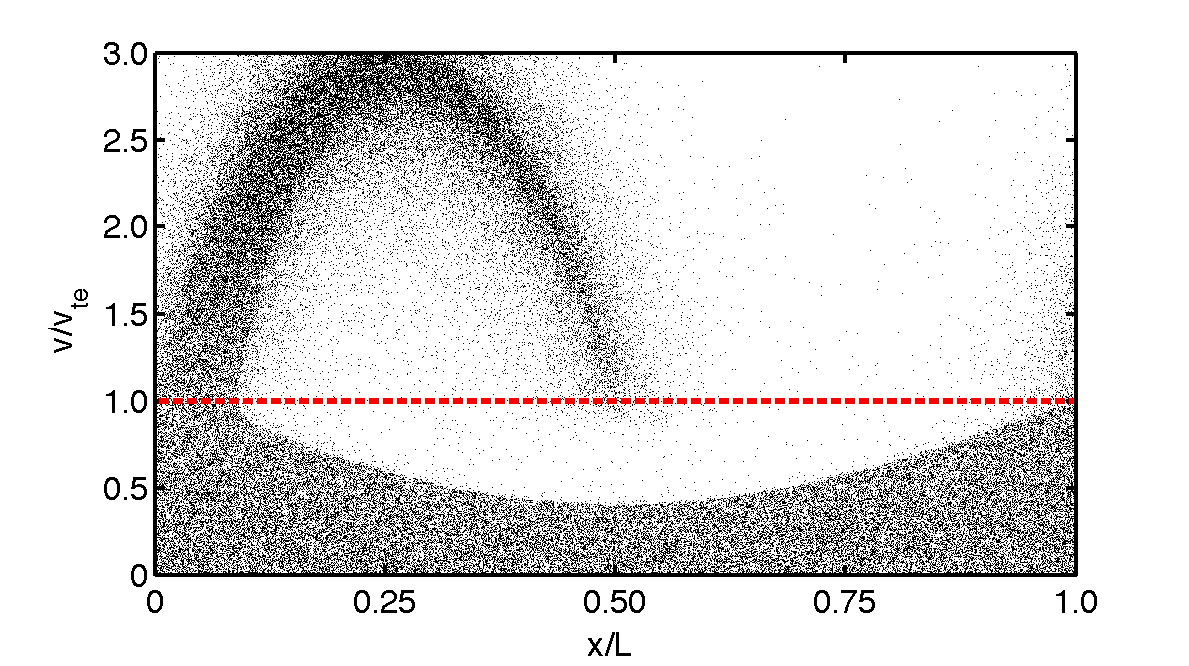}\vspace{-0.3in}\\
(c) $v_0=1.25v_\phi$ &(d) $v_0=v_\phi$
  \end{tabular}
   \caption{
   \label{fig:SLPIC}
   SLPIC accurately simulates electron trapping in phase space
   as long as $v_0$ is sufficiently greater than $v_\phi$.  
   The panels show SLPIC electrons in phase space, using the same physical parameters as Fig.~\ref{fig:resonantPIC}, for four cases with speed limits $v_0/v_{\phi}=$2, 1.5, 1.25, and 1, obtained with corresponding timesteps $\Delta t=\Delta x/v_0$; $v_0$ and $v_\phi$ are indicated by solid blue and dashed red lines, respectively.
For $v_0 \geq 1.5v_\phi$ (a, b), the SLPIC particles faithfully reproduce the PIC results; for $v_0=1.25v_\phi$ (c), the ``eye'' becomes distorted, especially at high velocities, and for $v_0=v_\phi$ (d), SLPIC completely fails to capture the physical wave-particle interaction.
}
\end{figure}

Next, we model the behavior of SLPIC particles in this traveling wave for various speed limits $v_{0}$, using the slowing-down function
in Eq.~(\ref{eq:heavisideBeta}).  We expect that when $v_0 \gg v_\phi$, the speed-limiting algorithm should still yield the correct electron distribution; indeed, this turns out to be the case.
In Fig.~\ref{fig:SLPIC}
we show the phase-space distribution of SLPIC particles at the same time as in Fig.~\ref{fig:resonantPIC}, with $v_0/v_\phi =$2, 1.5, 1.25, and $1$, obtained using timesteps of $\Delta t=\Delta x/v_0$ in each case.  For $v_0\geq 1.5v_\phi$, the SLPIC algorithm nicely preserves the phase-space structure within the separatrix of Fig.~\ref{fig:resonantPIC}; even for $v_0=1.5v_\phi$, where the speed limit occurs at the very edge of the separatrix, the salient features of the distribution do not differ appreciably from the PIC case. 
However, for $v_0=1.25v_\phi$, SLPIC speed-limits particles within the separatrix and begins to distort its shape in the higher-velocity portions of the phase space.

Figure~\ref{fig:SLPIC}(d) illustrates the disastrous results of setting the SLPIC speed limit exactly at the resonant wave phase velocity, $v_0=v_\phi$.  In this case, all particles with speeds above the phase velocity are slowed by the algorithm to travel at exactly the wave speed.
Due to the selection of Eq.~(\ref{eq:relativisticBeta}) as the speed-limiting function, acceleration or deceleration from the wave does not change the speed at which particles above the speed limit are moved through the simulation.  Particles above the speed limit thus ``see'' a constant wave phase and are continuously accelerated or decelerated accordingly.  Decelerated particles that drop below the speed limit can then physically respond to the wave, so that the upper-right region of Fig.~\ref{fig:SLPIC}(d) becomes increasingly depopulated.

\begin{figure}
   \centering
   \includegraphics[width=.90\textwidth]{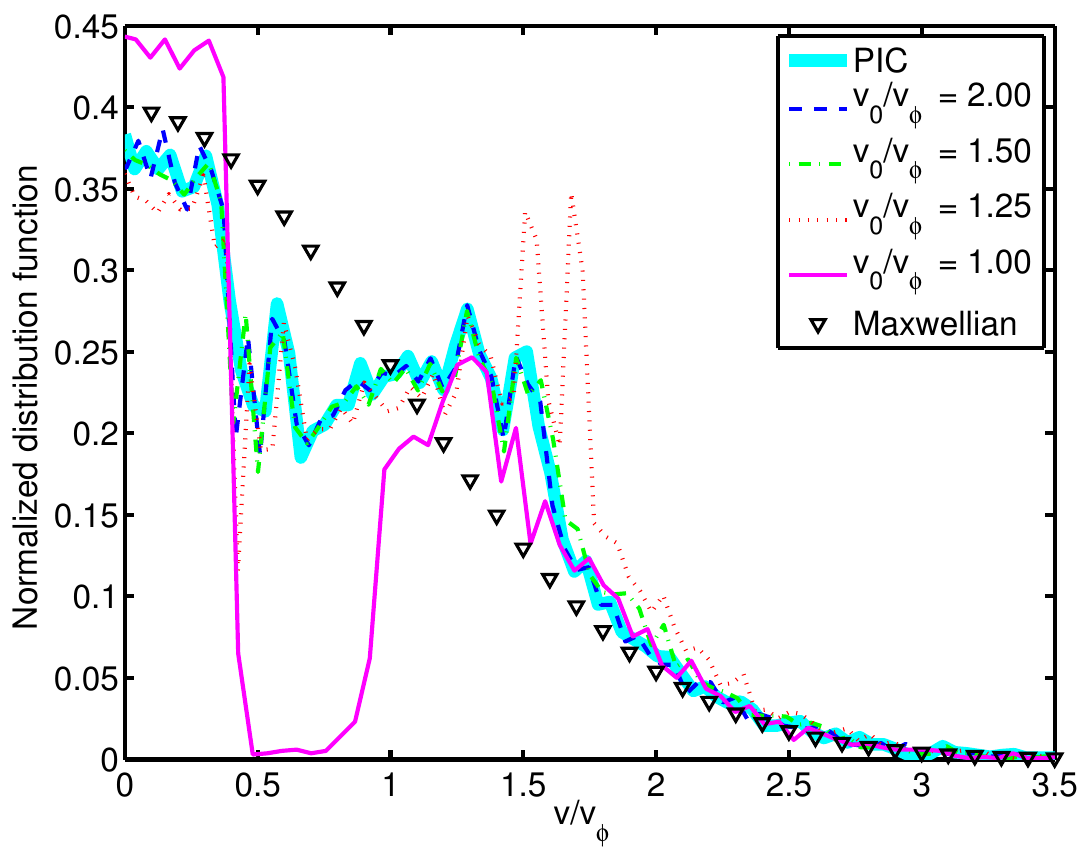}
   \caption{
   \label{fig:DistPlot}
Particle velocity distributions in the region bounded by the vertical green lines
in Fig.~\ref{fig:resonantPIC}, for the cases in Figs.~\ref{fig:resonantPIC} and \ref{fig:SLPIC}.  To highlight the quasilinear flattening behavior around the wave velocity $v_\phi$ (i.e., $0.5 \lesssim  v/v_{\phi} \lesssim 1.5$) arising from 
   wave-particle interactions, the initial Maxwellian particle distribution with thermal velocity $v_{te}=v_\phi$ is also shown.
As long as $v_{0}$ is above the flattened region ($v_0 \gtrsim 1.5 v_\phi$) where particles can be trapped, the SLPIC distributions
closely resemble the PIC distribution.  As $v_0$ approaches $v_\phi$,
the velocity distributions distort substantially.
}
\end{figure}

In Fig.~\ref{fig:DistPlot}, particle distributions for the cases in Figs.~\ref{fig:resonantPIC} and~\ref{fig:SLPIC} are shown for the region bounded by the vertical green lines in Fig.~\ref{fig:resonantPIC}.  The quasilinear flattening of the particle distribution function that arises from resonant wave-particle interactions is apparent; SLPIC introduces no substantive error (relative to the PIC distribution) so long as velocities in the flattened region are lower than the SLPIC speed limit.  We also see that, when this condition is violated, unphysical distortions of the distribution function ensue as portions of the trapped particle population experience unphysical wave-particle interaction due to speed limiting.  SLPIC is thus suitable for modeling wave-particle resonant interactions only if the speed limit $v_{0}$ is chosen to be well above the resonant phase velocities.

The simulations in this section were also repeated using a different slowing-down function, that of Eq.~(\ref{eq:relativisticBeta}).  For these cases, the steady-state particle distributions experienced distortion at high velocities [in the same manner shown in Fig. \ref{fig:SLPIC}(c)] even in the $v_{0}/v_\phi \simeq 2$ case.  Such behavior arises because this particular slowing-down function slows all particles in the distribution (even, mildly, those with $v \le  v_{0}$) and can thus interfere with the wave-particle resonant behavior.  Particles that should travel much faster than the wave can be slowed by the SLPIC algorithm to the wave velocity and thus respond strongly to the wave.
These results suggest that it may be prudent, when wave-particle interaction is of interest, to use slowing-down functions such as Eq.~(\ref{eq:heavisideBeta}) with little or no effect on particles below the speed limit.

We have thus illustrated the main effects of SLPIC on particle trajectories when particles are nearly resonant with a wave.  It is important to note, however, that we used 1D velocities.  The case with 3D velocities, especially for the slowing-down function Eq.~(\ref{eq:heavisideBeta}), is expected to be qualitatively similar after accounting for the possibility that a particle may be resonant with a wave when traveling at an angle to the wave front.  Capturing this interaction may require
higher speed limits than those that yielded good performance in this study, and for any advantageous speed limit, some fraction of particles traveling perpendicularly to the wave will not experience the correct wave-particle interaction.

\section{Summary}

Speed-limited particle-in-cell (SLPIC) simulation is a new technique
that allows kinetic PIC simulation with a larger timestep, by limiting the
maximum speed of particles (which otherwise follow the proper
trajectories).  The choice of speed limit controls the strength of
the approximation introduced by SLPIC: as the speed limit increases
beyond the speed of the fastest particle, SLPIC becomes identical to PIC.
Lowering the speed limit also lowers the plasma frequency, which allows
the timestep to be increased without risking the instability resulting in standard PIC from the failure to resolve the plasma frequency;
it also prevents particles from traveling too far during one (increased)
timestep.

SLPIC overcomes a common limitation of explicit PIC simulation:
even when the phenomena of
interest (hence fields and distribution functions) do not involve
plasma oscillations and change slowly compared
with the plasma frequency, the PIC timestep must be small enough to
resolve the plasma frequency to avoid instability.
Furthermore, SLPIC allows timesteps much larger than the
grid-cell-crossing time of the fastest particles, without actually allowing particles to cross multiple cells in one timestep.
SLPIC allows the timestep to be determined by the
phenomena of interest, rather than by the irrelevant and
much faster plasma oscillations.

SLPIC promises to be useful in cases where particle distribution functions
change slowly compared with the timestep required by PIC for stability
(and also for preventing particles from crossing too many cells in one timestep).
For example, SLPIC may be especially applicable to cases where the
electron distribution evolves on ion time scales.

In a typical SLPIC simulation with a cell size approximately equal to
some fraction of a Debye length, one might imagine setting the speed limit
some factor below typical electron speeds, but above physically-important
velocities such as the phase velocity of a plasma mode being simulated.
The timestep could then be increased by the same factor, speeding up
computation by roughly that factor, subject to some reduction
since the SLPIC particle-advance requires
more operations than standard PIC.  The extra operations are local
(to a single particle) and explicit, so they will take advantage of
memory cache and may be especially amenable to hardware acceleration;
relative to the potential SLPIC performance benefits,
they will not substantively affect overall scaling with problem size 
or computational resources.

The speed limit can in principle vary in both space and time, and this
offers intriguing possibilities for increasing the speed limit
(and decreasing the timestep) mid-simulation to increase and/or
verify the accuracy of approximation.

Modification of a PIC code to implement SLPIC is
localized to the integration of individual particle
trajectories and some considerations relating to particle weight (e.g., in charge
deposition).  Other aspects of PIC, such as collisions,
field solvers, boundary conditions, etc., should carry over to SLPIC
with little or no change.  SLPIC therefore promises a
particularly flexible and powerful
approach to increase the timestep of PIC simulations.

The basic SLPIC method has been tested in a 1D simulation of a
steady-state plasma sheath, successfully yielding an accurate potential profile
despite a timestep 320 times larger---and a running time 160 times shorter---than the corresponding PIC simulation.
This showed that SLPIC correctly captured the particle distributions on slow timescales.
In addition, we showed the behavior of 1D SLPIC test particles in a prescribed electrostatic wave, demonstrating that SLPIC correctly simulates the wave-particle interaction as long as the speed limit is sufficiently high that particles traveling (resonantly) with the wave are not speed-limited.  This latter point indicates that caution must be used with SLPIC when simulating phenomena such as Landau damping; SLPIC is unlikely to offer any speed-up over PIC for simulating wave-particle
interactions involving the fastest particles in a simulation, but could offer considerable advantage when the relevant wave-particle interactions involve slower particles.

\begin{acknowledgments}
This research was supported by U.S. Department of Energy SBIR Phase I/II Award DE-SC0015762; this material is based upon work supported by the National Science Foundation under Grant No.~PHY1707430. GW also received partial support from the Institute for Modeling
Plasma, Atmospheres, and Cosmic Dust (IMPACT) of NASA's Solar System
Virtual Institute (SSERVI).
\end{acknowledgments}

\appendix
\section{The speed-limited electrostatic particle-in-cell leapfrog algorithm}
In the electrostatic sheath example of Section~\ref{sec:sheath} we solve the following equations of motion of macroparticles in a self-consistent electric field $\mathbf{E}(\mathbf{x},t)$, with $\beta$ as a function of velocity magnitude only:
\begin{subequations}\label{eg:LimPtclDynAppendix}
   \begin{eqnarray} 
      \dot{\mathbf{x}}_{p}(t) &=& \beta\left(\left|\mathbf{v}_{p}(t)\right|\right) \mathbf{v}_{p}(t) \\
      \dot{\mathbf{v}}_{p}(t) &=& \beta\left(\left|\mathbf{v}_{p}(t)\right|\right) \frac{q_p}{m_p}\mathbf{E}(\mathbf{x}_p(t),t) \\
      \rho(\mathbf{x},t) &=& \sum_p w_p \beta\left(\left|\mathbf{v}_{p}(t)\right|\right) \delta[\mathbf{x} - \mathbf{x}_p(t)] \\
      \nabla \cdot \mathbf{E}(\mathbf{x},t) &=& \frac{\rho(\mathbf{x},t)}{\epsilon_0} \label{eg:LimPtclDynAppendixDotE} 
   \end{eqnarray}
\end{subequations}
Eqs.~(\ref{eg:LimPtclDynAppendix}) can be partially discretized in time via the leapfrog method by:
\begin{subequations}\label{eg:slpicIntegration}
   \begin{eqnarray} 
      \mathbf{v}_p^{n+\sfrac{1}{2}} &=& \mathbf{v}_p^{n-\sfrac{1}{2}} + \frac{q_p}{m_p}\mathbf{E}^n(\mathbf{x}_p^n)\int_{t^{n-\sfrac{1}{2}}}^{t^{n+\sfrac{1}{2}}}\beta\left(\left|\mathbf{v}_p(t) \right|\right) dt \label{eg:slpicIntegrationV}\\
      \mathbf{x}_p^{n+1} &=& \mathbf{x}_p^n + \beta\left(\left|\mathbf{v}_p^{n+\sfrac{1}{2}}\right|\right)\mathbf{v}_p^{n+\sfrac{1}{2}} \Delta t \\
      \rho^{n+1}(\mathbf{x}) &=& \sum_p w_p \beta\left(\left|\mathbf{v}_p^{n+\sfrac{1}{2}}\right|\right) \delta[\mathbf{x} - \mathbf{x}_{p}^{n+1}] \\
      \nabla \cdot \mathbf{E}^{n+1}(\mathbf{x}) &=& \frac{\rho^{n+1}(\mathbf{x})}{\epsilon_0} 
   \end{eqnarray}
\end{subequations}
where $\Delta t \equiv t^{n+1} - t^n$ is the timestep and $n$ is the timestep index.  Eq.~(\ref{eg:slpicIntegrationV}) is left in integral form due to the dependence of $\beta$ on the changing velocity.  
In our current implementation of SLPIC, which is meant to be simple and robust but not optimal, we evaluate the integral in Eq.~(\ref{eg:slpicIntegrationV}) using the modified midpoint method\cite{NumericalRecipes}, subdividing the timestep until the integral's value converges within 1\%.
We have performed the same simulation evaluating the integral to within $10^{-6}$\% error, with essentially identical results (e.g., the steady-state densities and potential are altered by less than 1\% when we use exactly the same particle injection); evaluating the integral to higher accuracy slows down the simulation without appreciably improving its accuracy.


\newcommand{\SortNoop}[1]{}

\end{document}